\documentclass[a4paper,11pt]{article}
\usepackage{pos}
\usepackage{float}

\title{Multi-messenger constraints on transient accelerators of ultra-high energy cosmic rays}

\title{Modeling gamma-ray signatures of particle acceleration
 in stellar clusters from GeV to PeV}

\author*[a]{A. Inventar}
\author[a]{S. Gabici}
\author[b]{E. Peretti}

\affiliation[a]{Université Paris Cité, CNRS, Astroparticule et Cosmologie \\ 
F-75013 Paris, France}

\affiliation[b]{INAF Osservatorio Astrofisico di Arcetri \\ Largo Enrico Fermi, 5, 50125, Firenze, Italy}

\emailAdd{inventar@apc.in2p3.fr}

\abstract{Young massive stellar clusters (YMSCs) have recently regained interest as PeVatron candidates, potentially accounting for the cosmic-ray (CR) knee as alternatives to isolated supernova remnants (SNRs). LHAASO's unique capability to detect photons above 0.1 PeV, hence tracing multi-PeV CRs, can provide critical constraints on galactic acceleration models when combined with H.E.S.S. and Fermi-LAT data. We investigate the transport of particles from YMSCs acceleration sites, namely wind termination shocks (WTS) or embedded SNRs, to nearby dense molecular clouds where proton-proton interactions produce high-energy gamma rays. We determine the necessary conditions, such as the distance between the acceleration site and the target, or the cluster's power and age, for detectable gamma-ray excesses and identify viable systems through parameter space exploration. By comparing with observations, we can constrain key physical parameters including WTS efficiency, diffusion coefficient and injection slope. Our analysis also examines whether some of LHAASO's unidentified sources might correspond to such cluster-cloud systems.}

\ConferenceLogo{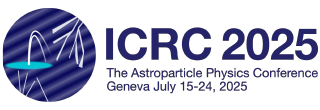}

\FullConference{39th International Cosmic Ray Conference (ICRC2025)\\
 15–24 July 2025\\
Geneva, Switzerland\\}

\begin{document}
\maketitle

\section{Introduction}

Young massive stellar clusters (YMSCs) are emerging as potential contributors to Galactic cosmic-ray (CR) acceleration, especially at high and very-high energies. Their dense populations of massive stars drive strong winds and supernovae, forming turbulent, magnetized stellar bubbles where acceleration can occur at wind termination shocks (WTS) and embedded supernova remnant (SNR) shocks \citep{Gabici2023}.

Gamma-ray observations have revealed extended emission above 100TeV near clusters like Cygnus Cocoon and Westerlund 1 \citep{LHAASOCyg}\citep{Aharonian2022}, indicating the presence of PeVatrons. These results suggest that YMSCs may complement isolated SNRs as sources of Galactic CRs.

Sustained acceleration in YMSCs can be done due to prolonged energy input and efficient particle confinement. Hadronic interactions with nearby molecular clouds can produce observable gamma rays, offering indirect evidence of the CR population. \citep{Gabici2009}

This work investigates whether cluster–cloud systems can account for gamma-ray emission up to hundreds of TeV. By comparing model predictions with observations, we constrain key parameters such as CR efficiency, diffusion, and spectral properties, and assess whether some of LHAASO’s unidentified sources may arise from such interactions.

\section{Transport of particles and maximal distances for CR excess}

We will consider accelerated particles escaping a low density stellar bubble, and being transported diffusively in the interstellar medium (ISM) up to some target molecular cloud where pp interactions  take place, creating gamma-rays. This model can therefore be applied only for distances bigger than the stellar bubble's radius $R_b$ (otherwise advection should also be taken into account, as done in \citep{Morlino2021}). $R_b$ is given by \citep{Weaver1977}, as computed in \citep{Gabici2023}: 

\begin{align} 
 \label{Rb}
    & R_b=260 ~  \biggl(\frac{\eta N_*}{100}\biggr)^{1/5} \biggl(\frac{n_0}{~\mathrm{cm^{-3}}}\biggr)^{-1/5} \biggl(\frac{t}{10~\mathrm{Myr}}\biggr)^{3/5} \mathrm{pc}
\end{align}

with $N_*$ the number of massive stars (between 8 and 150 $M_{\odot}$) in the cluster, $n_0$ the ISM density, $u_w$ the wind speed, t the age of the cluster and $\eta$ a correction factor due to radiative losses. Typically, for powerful clusters around dense regions, $N_*=100$, $n_0 \sim 100  ~\mathrm{cm}^{-3}$, $u_w \sim 3000 ~\mathrm{km/s}$, $t=5 ~\mathrm{Myr}$, $\eta=0.22$ \citep{Vieu2022}, which gives $R_b\sim 50 $pc. \\

\subsection{Impulsive injection}

We begin by analysing an impulsive injection of cosmic‑ray protons, both spatially and temporally localised, as expected for a supernova remnant exploding inside a stellar cluster while ignoring collective effects.
The release is assumed to occur at $R = 0$ and $t = 0$, injecting a total CR energy $W_{CR}$.
With a differential spectrum $Q(E)$, the injected energy satisfies

\begin{equation}
\label{eq:Wcr}
W_{CR} = \int_{E_{min}}^{E_{max}} {\rm d}E Q(E)E .
\end{equation}

Limiting ourselves to relativistic protons ($E_{min}=m_pc^2$) with a maximal energy $E_{max}=3 \rm ~ PeV$ and a power‑law spectrum, we have

\begin{equation}
Q(E)=Q_{min}\left(\frac{E}{E_{min}}\right)^{-\alpha} \exp \left({-\frac{E}{E_{max}}} \right ),
\end{equation}

and for $\alpha>2$, this simplifies to

\begin{equation}
\label{eq:inji}
Q(E)\simeq(\alpha-2),\frac{W_{CR}}{(m_pc^{2})^{2}}\left(\frac{E}{m_pc^{2}}\right)^{-\alpha} \exp \left({-\frac{E}{E_{max}}} \right ).
\end{equation}

After injection, CRs diffuse through isotropic turbulence with $D(E)=D_{10}[E/(10 ~m_pc^{2})]^{\delta}$. Noting $ \tau_{d}=\frac{R^2}{6D(E)}$ the typical diffusion timescale and $\tau_{pp} \sim 6\cdot10^7 (\frac{n_0}{1cm^{-3}})^{-1} \rm yr$ the proton lifetime \citep{Aharonian1996}, we obtain that for typical parameters ($D_{10}\sim 10^{28} \rm cm^2 s^{-1}, R\sim 100 \rm pc$), $\tau_{pp} \gtrsim \tau_{d}$. Therefore the energy losses can be neglected and the distribution of CRs obeys:

\begin{equation}
\label{eq:transport}
\frac{\partial n}{\partial t}= \frac{1}{R^{2}}\frac{\partial}{\partial R}\left(D R^{2}\frac{\partial n}{\partial R}\right) + \frac{Q(E)}{4\pi R^{2}}\delta(t)\delta(R),\end{equation}

whose well known solution is \citep{Aharonian1996}

\begin{equation}\label{snr}
n(t,E,R)=\frac{Q(E)}{\pi^{3/2}R_{d}^{3}}\exp\left[-\left(\frac{R}{R_{d}}\right)^{2}\right],
\end{equation}

with $R_{d}(t,E)=\sqrt{4D(E)t}$. For $R\ll R_{d}$ the distribution is flat; for $R\gg R_{d}$ it declines rapidly.

For order‑of‑magnitude estimates we take $n\approx Q/[(4\pi/3)R_{d}^{3}]$ inside $R_{d}$ and zero outside, and we also drop the cutoff term $\exp \left({-\frac{E}{E_{max}}} \right )$. Approximating the local CR spectrum by $E^{2}n_{LIS}\simeq0.6 (E/m_pc^{2})^{-0.7}$eVcm$^{-3}$ \citep{Gabici2022}, an excess over the Galactic background persists up to

\begin{equation}
t_{max} \approx 7 \times 10^4 \left( \alpha - 2 \right)^{\frac{2}{3}} \left( \frac{W_{CR}}{10^{50}{\rm erg}} \right)^{\frac{2}{3}} \left( \frac{10^{28}~{\rm cm^2/s}}{D_{10}} \right) \left( \frac{E}{m_p c^2} \right)^{1.8-\frac{2}{3}\alpha-\delta} \rm yr
\end{equation}

within a radius

\begin{equation}
R_d(t_{max}) \approx 10^2 \left( \alpha - 2 \right)^{\frac{1}{3}} \left( \frac{W_{CR}}{10^{50}{\rm erg}} \right)^{\frac{1}{3}} \left( \frac{E}{m_p c^2} \right)^{0.9-\frac{1}{3} \alpha} \rm pc
\end{equation}
from the accelerator.

For typical parameters, such as $\alpha=2.2$ and $\delta=0.4$, this yields to $t_{max}\sim2.5\times10^{4}$yr and $R_{d}\sim80$pc for 10GeV CRs, and $t_{max}\sim8\times10^{3}$yr and $R_{d}\sim0.6$kpc for 1PeV CRs.
Outside these limits the source has little influence on the ambient CR density. We also note that the dependence in  $\alpha$ and $\delta$ can have drastic changes, especially at higher energies.
An excess factor $\Delta=n/n_{LIS}$ is found at $t\approx\Delta^{-2/3}t_{max}$ and $R\approx\Delta^{-1/3}R_{d}(t_{max})$. \\

Finally, the diffusion regime is valid only when $\Delta t>\dfrac{L}{c}$, with $\Delta t$ the time passed since the injection of the considered CRs, and $L=\frac{3D(E)}{c}$ the mean free path of the CRs. When this is not the case, we enter the ballistic regime. Typically, for $D_{10}=10^{28} \mathrm{cm}^2 \mathrm{s}^{-1}$, $\delta=0.4$ and $E=1 \mathrm{PeV}$, this gives $\Delta t \gtrsim 160$ yrs, and in terms of distance, it translates in $R> \dfrac{3D}{c}=L \sim 33$pc. This means that the diffusion regime is valid for distances bigger than this value, which will be always the case here since $R_b$ is already bigger.

\subsection{Continuous injection}

This approach can be extended to sources with continuous CR injection beginning at $t = 0$. For modeling a steady CR injection rate $\dot{W}_{CR}$, the final term in Eq.~\ref{eq:transport} should be replaced by $\dot{Q}/(4 \pi R^2) \delta(R)$, where $\dot{Q}$ follows Eq.~\ref{eq:inji} with the replacements $Q \rightarrow \dot{Q}$ and $W_{CR} \rightarrow \dot{W}_{CR}$. The transport equation then yields \citep{Aharonian1996} :

\begin{equation}
\label{eq:continuous}
n(t,E,R) = \frac{\dot{Q}(E)}{4 \pi D(E) R} \mathrm{erfc} \left( \frac{R}{R_d(t,E)} \right) ~ ,
\end{equation}

\vspace{2mm}
where $\mathrm{erfc}$ denotes the complementary error function.

Following the impulsive injection case, we approximate Eq.~\ref{eq:continuous} as $n \sim \dot{Q}/4 \pi D R$ for $R < a \times R_d$ and zero otherwise, with $a = 1/\sqrt{2}$ ensuring CR particle number conservation. This simplification predicts a CR excess over Galactic background within a radius:

\begin{equation}
R_{max} \approx 3 \times 10^2 \left(\alpha - 2 \right) \left( \frac{\dot{W}_{CR}}{10^{37} {\rm erg/s}} \right) \left( \frac{10^{28}{\rm cm^2/s}}{D_{10}} \right) \left( \frac{E}{m_p c^2} \right)^{2.7-\alpha-\delta} \rm pc.
\end{equation}

\vspace{2mm}

Here, $\dot{W}_{CR}$ is normalized to $\sim$10 \% of a powerful star cluster's mechanical power (see e.g\citep{Menchiari2024}). For $\alpha = 2.2$ and $\delta = 0.4$, this yields to $R_{max} \approx 60$~pc at 10 GeV and  $R_{max} \approx 0.2$~kpc at 1 PeV.  We note again that a small change in  $\alpha$ and $\delta$ can have drastic changes, but that for $\alpha = 2.2$ and $\delta = 0.5$, the maximal distance becomes energy independent. At distances $R < R_{max}$, a CR excess $\Delta \sim R_{max}/R$ appears for times exceeding:

\begin{equation}
t_{min} \approx 10^6 \frac{\left(\alpha - 2 \right)^2}{\Delta^2} \left( \frac{\dot{W}_{CR}}{10^{37} {\rm erg/s}} \right)^2 \left( \frac{10^{28}{\rm cm^2/s}}{D_{10}} \right)^3 \left( \frac{E}{m_p c^2} \right)^{5.4-2 \alpha-3 \delta} \rm yr.
\end{equation}

\vspace{2mm}

For $\alpha = 2.2$ and $\delta = 0.4$, this reduces to $t_{min}$ of the order of $\sim$ few kyr, significantly shorter than typical cluster lifetimes ($\sim$Myr). Only a substantially reduced $D_{10}$ could make $t_{min}$ comparable to cluster lifetimes, making $t_{min} \approx 0$ a reasonable approximation in most cases. 

\section{\textbf{Excess in gamma-rays}}

From the proton spectra given in Eqs.~\ref{snr} and \ref{eq:continuous}, we can compute the $\gamma$-ray emissivity esulting from proton-proton (p-p) interactions between the accelerated cosmic-ray protons and the ambient protons within a molecular cloud of mass $M_{\rm cloud}$, density $n_H$, and located at a distance $d$ from Earth. Using the parametrization provided by \citep{Kafexhiu2014}, this emissivity can be converted into a $\gamma$-ray flux to obtain:

\vspace{-2mm}

\begin{equation}
    \phi_{\gamma} = \int \frac{d\sigma}{dE_\gamma} (E, E_\gamma)\, n(E)\, dE \frac{c}{4 \pi d^2} \frac{M_{\rm cloud}}{m_p}
\end{equation}

Similarly, assuming the local interstellar spectrum (LIS) of cosmic rays follows $E^2 n_{\rm LIS} \simeq 0.6\, (E/m_p c^2)^{-0.7}$~eV\,cm$^{-3}$, we can estimate the $\gamma$-ray flux resulting from the interaction of the CR sea with a molecular cloud.

\vspace{2mm}

To compute the total $\gamma$-ray background along the line of sight, one must also account for interactions with diffuse gas. This is done by applying a multiplicative correction factor of $(n_{\rm col}^C + n_{\rm col}^D)/n_{\rm col}^C$, where $n_{\rm col}^C$ is the column density of the molecular cloud and $n_{\rm col}^D$ is the column density of the diffuse interstellar gas in the observed region.

\vspace{2mm}

We also take into consideration the point-source sensitivities of several $\gamma$-ray instruments, including Fermi-LAT, LHAASO, and CTA. The sensitivity to extended sources is obtained by scaling the point-source sensitivity by a factor $\theta_{\rm cloud}/\theta_{\rm res}(E)$, provided this ratio exceeds unity. Here, $\theta_{\rm cloud}$ denotes the angular size of the cloud, and $\theta_{\rm res}(E)$ is the energy-dependent angular resolution of the detector.

\section{Parameter space constraints for a $\gamma$-ray excess}

To extend the previous analytic considerations on $R_{max}$, we can now plot the $\gamma$-ray flux arising from p-p interactions against the distance between the cluster and the cloud, at a given photon energy (taken as 0.3 PeV here) and for several sets of parameters. We consider here a typical powerful cluster with $ \dot{W}_{CR}= 3 \cdot 10^{37} ~ \rm erg.s^{-1}$ ( \citep{Menchiari2024}) and supernovae with $W_{CR}=10^{50} \rm erg$. Moreover, we take as an example a molecular cloud of $10^5M_{\odot}$ and located at 2kpc from Earth. 

\begin{figure}[H]
 \centering
 \includegraphics[width=0.49\textwidth]{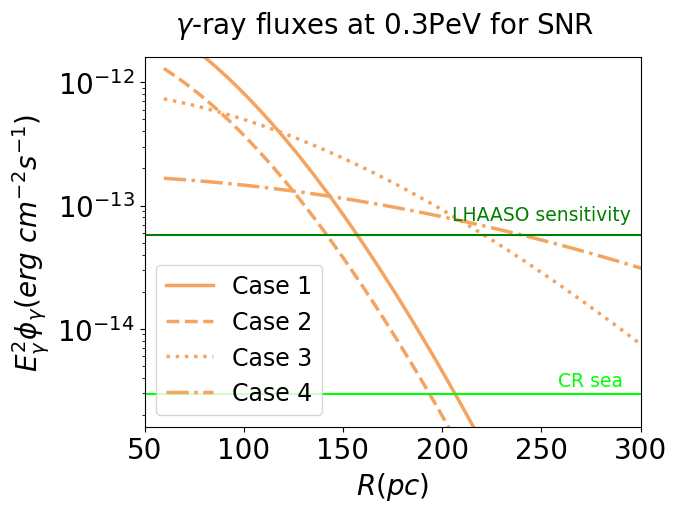}     
 \includegraphics[width=0.49\textwidth]{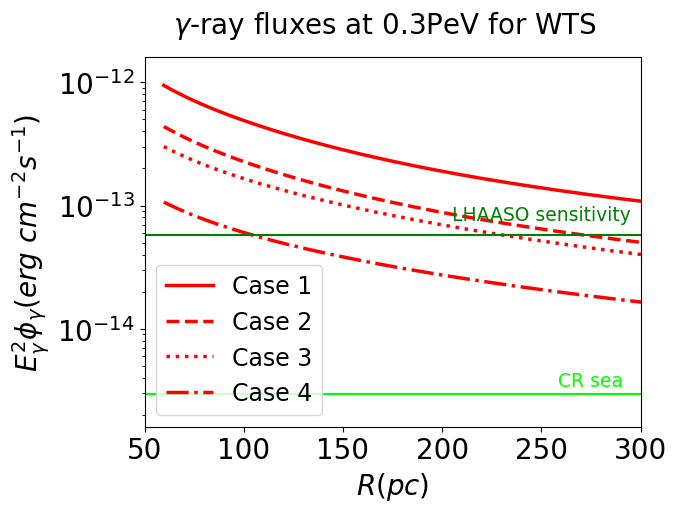}      
  \caption{ {\bf{Left:}} Gamma-ray fluxes at 0.3 PeV from impulsive injection: embedded SNRs, $W_{CR}=10^{50} \rm erg$ and age fixed at $t=10^4 \rm years$ (parameter degenerate with $D_{10})$. {\bf{Right:}} Gamma-ray fluxes at 0.3 PeV from continuous injection: WTS, $ \dot{W}_{CR}= 3 \cdot 10^{37} ~ \rm erg.s^{-1}$. For both plots, the target is a molecular cloud of mass $M_{cloud}=10^5M_{\odot}$ and at d=2kpc from us. In light green is plotted the gamma-ray background from the CR sea interacting with the cloud and in dark green is the LHAASO point-like sensitivity. Case 1,2,3,4 refer respectively to the parameter set 1) $\delta=0.3$, $\alpha=2.0$, $D_{10}=10^{27} \mathrm{cm}^2 \mathrm{s}^{-1}$; 2) $\delta=0.3$, $\alpha=2.1$, $D_{10}=10^{27} \mathrm{cm}^2 \mathrm{s}^{-1}$; 3) $\delta=0.4$, $\alpha=2.0$, $D_{10}=10^{27} \mathrm{cm}^2 \mathrm{s}^{-1}$; 4) $\delta=0.3$, $\alpha=2.0$, $D_{10}=10^{28} \mathrm{cm}^2 \mathrm{s}^{-1} $.}
  \label{phi_vs_R}
\end{figure}

\vspace{2mm}
On Fig.\ref{phi_vs_R}, we show the influence of a change in the acceleration and transport parameters at this very high energy ($\sim 300 ~ \rm TeV$). In particular, we remark that there is a whole band of the parameter space (between the two green lines) where there can be an excess, such as computed in the analytic part, but where it cannot be detected (by LHAASO here). Therefore, in order to obtain a detectable excess of gamma-rays at these energies, the parameter space is reduced to small values of $\delta$, $\alpha$ and $D_{10}$, implying a hard injection, a near-Kolmogorov diffusion and a suppression of diffusion coefficient typically via streaming instabilities \citep{Bell2004}. We also see that the most optimistic cases allow an effective maximal distance for detectable excess of $R_{max, ~d} \sim 300 \rm~ pc$. \\

\begin{figure}[H]
 \centering
 \includegraphics[width=0.593\textwidth]{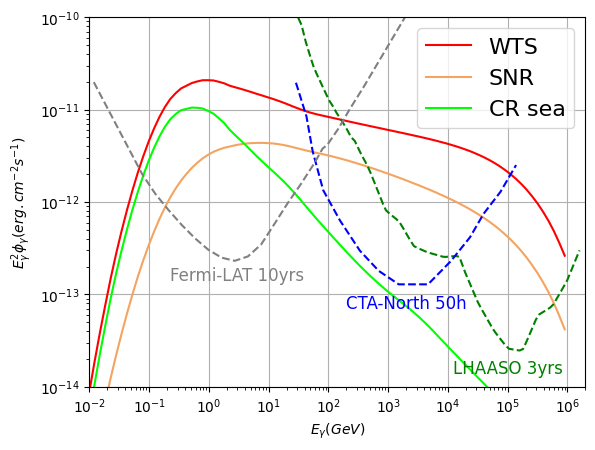}    
  \caption{$\gamma$-ray spectra from a molecular cloud of mass $M_{cloud}=10^5M_{\odot}$, 60pc away from the cluster and at d=2kpc from us. Incoming CRs are accelerated in embedded SNRs ($W_{CR}=10^{50} \rm erg$ and $t=10^5 \rm years$), or in WTS ($ \dot{W}_{CR}= 3 \cdot 10^{37} ~ \rm erg.s^{-1}$) or are the CR sea. Parameters are fixed to $\delta=0.3$, $\alpha=2.0$, $D_{10}=10^{27} \mathrm{cm}^2 \mathrm{s}^{-1} $. Point-like differential sensitivity of LHAASO, CTA North and Fermi-LAT are shown in dashed lines.}
  \label{phi_vs_E}
\end{figure}

Naturally, at lower energies (GeV), the background of CRs is much bigger and we do not have this phenomenon of effective maximal distances brought by the detectors' sensitivities. To explore what happens between these extreme cases, we plot now in Fig.\ref{phi_vs_E} the  $\gamma$-ray fluxes as function of energy, fixing a given distance of 60 pc. We consider the same cloud than before and the same classes of sources (CR sea, WTS, embedded SNRs). We also show the point-like differential sensitivity of several  $\gamma$-ray detectors (LHAASO,CTA North, Fermi-LAT).

This plot first shows that we start having effective maximal distances from $E_{\gamma} \sim 1 \rm ~ TeV$. Then, for the considered set of parameters, the wind gives a bigger flux than the SNRs in the overall energy range. However, this highly depends on the parameters and cannot be interpreted as a general behavior. By combining space and energy dependences shown in Fig.\ref{phi_vs_R} and Fig.\ref{phi_vs_E}, we are able to find the configurations of parameters enabling a detectable $\gamma$-ray excess in the desired energy range, and the corresponding systems star cluster/cloud. A full parameter space scan will be made in a paper in preparation. The next step is to find a $\gamma$-ray source that can correspond to this physical situation, and compare models to observed $\gamma$-ray flux, in order to infer parameters values. 

\section{Application to W43 region}

W43 is a very rich and extended star forming region \citep{Motte2003}. It contains a powerful star cluster towards its center, called W43 main. Its estimated distance of $d=5.5$kpc \citep{Zhang2014} while its age is estimated at 5-6 Myr \citep{Bally2010} even though there is no precise consensus on it. Several giant molecular clouds up to $10^6 M_{\odot}$ are also present in the vicinity of this cluster \citep{Miville2017}. In addition, its far-infrared luminosity is estimated at $L_{F-IR}\sim 4\times 10^6 L_{\odot}\sim 1.5 \times 10^{40}$erg/s \citep{Smith1978}, from which we can estimate roughly a bolometric luminosity $L_{bol}\sim 3L_{IR} \sim 4\times 10^{40} $erg/s. Using star cluster population simulation as in \citep{Vieu2022} and mass-luminosity relation \citep{Menchiari2023}, it is possible to infer that the upper limit value of the wind luminosity of this cluster is $L_{w}\sim 3 \times 10^{38} \rm erg/s$. Noting $\epsilon_w$ the WTS efficiency, this implies $\dot{W}_{CR}=~ 3 \epsilon_w \times 10^{38} \rm~erg/s$. We will use these values in the following. 

\begin{figure}[H]
 \centering
 \includegraphics[width=0.4\textwidth]{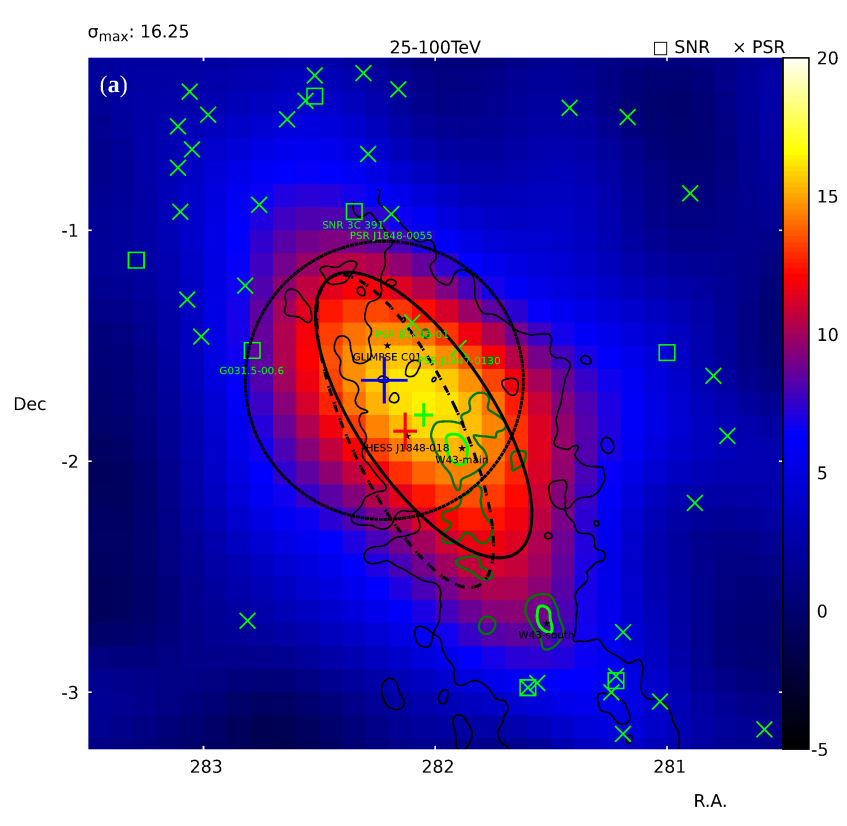}     
 \includegraphics[width=0.51\textwidth]{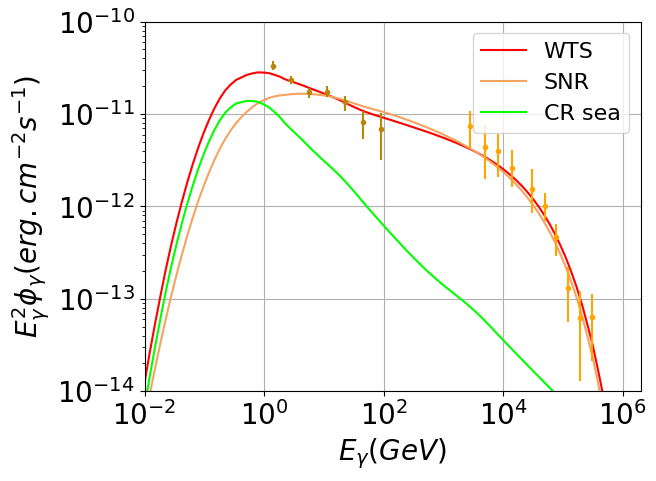}      
  \caption{ {\bf{Left:}} $\gamma$-ray significiance map of LHAASO J1848-0153u between 25 and 100 TeV, with star cluster W43 main in the center. {\bf{Right}:} $\gamma$-ray spectra fitted to LHAASO J1848-0153u and the corresponding Fermi datapoints derived in \citep{Yang2020}. Emissions are computed for a molecular cloud of mass $M_{cloud}=10^6M_{\odot}$ at d=5.5kpc from us, where incoming CRs are accelerated in embedded SNRs or in WTS. For WTS, the parameters used are $\dot{W}_{CR}=~ 3 \epsilon_w \times 10^{38} \rm~erg/s$,  $R=75\rm ~pc$, $\delta=0.33$, $\alpha=2.0$, $D_{10}=10^{27} \mathrm{cm}^2 \mathrm{s}^{-1} $ while for SNR: $W_{CR}=2\times 10^{50} \rm erg$, $t=3 \times 10^4 ~ \rm yrs$, $R=55\rm ~pc$,  $\delta=0.3$, $\alpha=2.0$, $D_{10}=3 \times 10^{27} \mathrm{cm}^2 \mathrm{s}^{-1} $.}
  \label{w43}
\end{figure}

Fig.\ref{w43} shows this region as recently seen in $\gamma$-rays by LHAASO \citep{LHAASO2024}. This source has also been seen by HESS \citep{HESS2018}. W43 main is located towards the center of the $\gamma$-ray emission, and there are almost no isolated SNRs or pulsars counterpart in the region of interesest. Therefore, we will fit our models to this data and the Fermi data computed in \citep{Yang2020} to infer acceleration and transport parameters.

First, we see that the emission is not shell-like (conversely to Wd 1 for instance \citep{Harer2023}). Moreover, the half-size of the ellipse
corresponds to $\sim 50 ~ \rm pc$, while electrons are expected to be cooled on smaller distances at these very high energies. Therefore, these elements disfavor a leptonic scenario. As for hadronic scenarios, we plot in the right part of Fig\ref{w43} the fits from the models studied before. For both models, a cutoff at $\sim 0.4 ~\rm PeV$ is needed, which is quite below the proton knee (3PeV) but is coherent with the Hillas criterium for $B \sim 10 \mu G$. We see that the best fit with WTS is better for low energy than with one SNR, and that it requires a distance which is more flexible (75pc instead of 55). Of course, a fit with several SNRs of different ages would give even better results but this would add too many free parameters. Therefore, the WTS scenario is favored in this case. Thus, we can estimate that the injection should be near $\alpha=2.0$, the  diffusion should be near Kolmogorov regime and we can constrain the following ratio as $\epsilon \frac{10^{28} \mathrm{cm^2s^{-1}}}{D_{10}} \sim 1$.

\section{Conclusion}

In this work, we modeled the transport of CRs initially accelerated in a star cluster (whether from WTS or embedded SNR shock) up to a molecular cloud where  $\gamma$-rays can be created via p-p interactions. We studied the parameter space in which an excess of  $\gamma$-rays can be seen on Earth, especially the space and energy  dependences. We noted that the WTS presents a larger part of the parameter space that can be of interest for LHAASO, with respect to the SNR scenario. A full parameter space scan will be done in a paper in preparation.

We then applied these models on W43, a system star cluster/giant molecular cloud with actual $\gamma$-ray data. First excluding the leptonic scenario because of the morphology, we then infered which parameters of the model would be needed in order to fit the datapoints. We concluded that the WTS scenario was more able to fit data and we constrained in particular the ratio between WTS efficiency and diffusion coefficient as  $\epsilon \frac{10^{28} \mathrm{cm^2s^{-1}}}{D_{10}} \sim 1$. 

Future work should consider more such systems, to make the estimation of the parameters more accurate. Moreover, scenarios where the $\gamma$-rays are produced within the stellar bubble (target being dense clumps) shoud also be studied in the future.

\end{document}